\documentclass{PoS}
\rightline{\parbox{4cm}{\large\rm
Edinburgh 2008/48\\
SHEP-0835
}}

\usepackage{amsmath}
\usepackage{amssymb}
\usepackage{psfrag}
\usepackage{array}
\usepackage{graphicx}
\usepackage{wrapfig}
\usepackage{axodraw}
\usepackage{cite}

\newcommand{\be}{\begin{equation}}
\newcommand{\ee}{\end{equation}}
\newcommand{\bc}{\begin{center}}
\newcommand{\ec}{\end{center}}

\def\gev{\,\mathrm{GeV}}
\def\mev{\,\mathrm{MeV}}

\def\SU{\mathrm{SU}}

\def\rpisq{\langle r_\pi^2\rangle}
\def\rpisqsu2sim{0.354(31)}  

\def\rpisqphyslong{0.418(31)}

\title{$K_{l3}$ and pion form factors using partially twisted boundary conditions}

\ShortTitle{$K_{l3}$ and pion form factors}

\author{J.M. Flynn, H. Pedroso de Lima, C.T. Sachrajda\\
        School of Physics and Astronomy, University of Southampton,
        Southampton SO17 1BJ, UK}

\author{P.A. Boyle, C. Kelly, C.M. Maynard, J.M. Zanotti\\
        School of Physics and Astronomy, University of Edinburgh,
        Edinburgh EH9 3JZ, UK}

\author{A. J\"uttner\\
        Institut f\"ur Kernphysik, Johannes-Gutenberg Universit\"at Mainz,
        D--55099 Mainz, Germany}

\author{RBC and UKQCD Collaborations}

\abstract{We compute the $K_{\ell 3}$ and pion form factors using
  partially twisted boundary conditions. The twists are chosen so that
  the $K_{\ell 3}$ form factors are calculated directly at zero momentum
  transfer $(q^2=0)$, removing the need for a $q^2$ interpolation,
  while the pion form factor is determined at values of $q^2$ close to
  $q^2=0$.
  The simulations are performed on an ensemble of the RBC/UKQCD
  collaboration's gauge configurations with Domain Wall Fermions and
  the Iwaski gauge action with an inverse lattice spacing of
  1.73(3)\,GeV.
  Simulating at a single pion mass of 330\,MeV, we find the pion
  charge radius to be $\langle r^2\rangle_{\rm
    330\,MeV}=0.354(31)\,{\rm fm}^2$ which, using NLO SU(2) chiral
  perturbation theory, translates to a value of $\langle
  r_\pi^2\rangle=0.418(31)\,{\rm fm}^2$ for a physical pion.
  For the value of the $K_{\ell 3}$ form factor, $f_{K\pi}^+(q^2)$,
  determined directly at $q^2=0$, we find a value of
  $f_{K\pi}^+(0)=0.9742(41)$ at this particular quark mass, which agrees well
  with our earlier result (0.9774(35)) obtained using the standard,
  indirect method.}

\FullConference{The XXVI International Symposium on Lattice Field Theory\\
		 July 14-19 2008\\
		 Williamsburg, Virginia, USA}

\begin{document}

%
\section{Introduction}
%

Over the last two years as part of our Domain Wall Fermion (DWF)
physics programme we have been looking at the $K\rightarrow \pi \ell
\nu_\ell$ ($K_{\ell 3}$) form factor at zero momentum transfer.
Since the experimental rate for $K_{\ell 3}$ decays is proportional to
$|V_{us}|^2 |f_{K\pi}^+(0)|^2$, a lattice calculation of the form factor,
$f_{K\pi}^+(q^2)$ at $q^2=0$, provides an excellent avenue for the
determination of the Cabibbo-Kobayashi-Maskawa (CKM)
\cite{Cabibbo:1963yz} quark mixing matrix element, $|V_{us}|$.

\begin{wrapfigure}{r}{0.45\textwidth}
\bc
\includegraphics[width=0.45\textwidth]{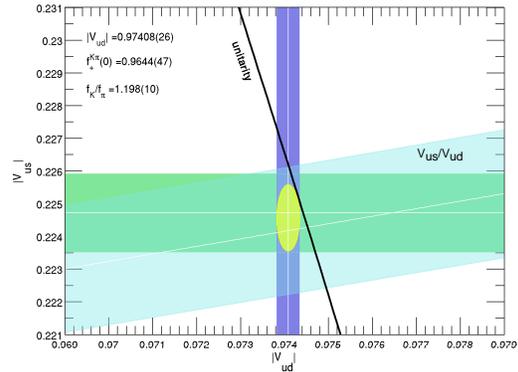}
\caption{Bands showing the current limits on
  $|V_{ud}|$ \cite{Yao:2006px}, and $|V_{us}|$\cite{Boyle:2007qe}.}
\label{fig:vusvud}
\ec
\end{wrapfigure}

The uncertainty in the unitarity relation of the CKM matrix 
$
\left|V_{ud}\right|^2 + \left|V_{us}\right|^2 = 1 
$
(we ignore $\left|V_{ub}\right|$ since this is very small), is
dominated by the precision of $\left|V_{us}\right|$.
In Fig.~\ref{fig:vusvud} we show the latest determinations of
$|V_{ud}|$ \cite{Yao:2006px} and $\left|V_{us}\right|$
\cite{Boyle:2007qe}.
For comparison, we also show the unitarity relation.
Since it is important to establish unitarity with the best precision
possible, it is essential that we decrease the error in $|V_{us}|$.

The value of $f_{K\pi}^+(0)$ used in determining $\left|V_{us}\right|$ in
figure~\ref{fig:vusvud} was determined using standard methods
\cite{Becirevic:2004ya,Dawson:2006qc} involving periodic boundary
conditions in the recent paper \cite{Boyle:2007qe}.
There, the $K_{\ell 3}$ form factor is calculated at $q^2_{max}=(m_K -
m_{\pi})^2$ and several negative values of $q^2$ for a variety of
quark masses.
This allows for an interpolation of the results to $q^2=0$. 
The form factor is then chirally extrapolated to the physical pion and
kaon masses.
The final result for $f_{K\pi}^+(0)$ quoted is then \cite{Boyle:2007qe}
$
f_{K\pi}^+(0) = 0.9644(33)(34)(14)
$
where the first error is statistical, and the second and third are
estimates of the systematic errors due to the choice of
parametrisation for the interpolation and lattice artefacts,
respectively.
This gives us a value of $\left|V_{us}\right|=0.2249(14)$.

More recently, we have developed a method that uses partially twisted
boundary conditions to calculate the $K_{\ell 3}$ form factor directly
at $q^2=0$ \cite{Boyle:2007wg}, thereby removing the systematic error
due to the choice of parametrisation for the interpolation in $q^2$.
We have also used partially twisted bc's to calculate the pion form
factor at values of $q^2$ below the minimum value obtainable with
periodic bc's.
In contrast to recent studies this allows for a direct evaluation of
the charge radius of the pion.
The method was developed and tested in \cite{Boyle:2007wg} and now
applied in a simulation with parameters much closer to the physical
point.

In this paper we discuss our findings for the pion form factor from
\cite{Boyle:2008yd} and our progress in improving the precision of our
result for $f_{K\pi}^+(0)$ from \cite{Boyle:2007qe} using partially twisted
boundary conditions.

%
\section{Simulation Parameters}
%

The computations are performed using an ensemble with light quark
mass $am_u=am_d=0.005$ and strange quark mass $am_s=0.04$ from a set
of $N_f=2+1$ flavour DWF configurations with $(L/a)^3\times T/a\times
L_s=24^3\times 64\times 16$ which were jointly generated by the
UKQCD/RBC collaborations \cite{Allton:2008pn} using the QCDOC
computer.
The gauge configurations were generated with the Iwasaki gauge action
with an inverse lattice spacing of $a^{-1}=1.729(28)\mathrm{GeV}$. The
resulting pion and kaon masses are $m_\pi \approx 330\mathrm{MeV}$ and
$m_K \approx 575\mathrm{MeV}$, respectively.

In this work we use single time-slice stochastic sources
\cite{Boyle:2008rh}, for which the elements of the source are randomly
drawn from a distribution $\mathcal{D}=\mathbb{Z}(2)\otimes
\mathbb{Z}(2)$ which contains random $\mathbb{Z}(2)$ numbers in both
its real and imaginary parts.
With sources of this form we find that the computational cost of
calculating quark propagators is reduced by a factor of 12.
For more details on the simulations, see \cite{Boyle:2008yd}.

%
\section{The Form Factors}
\label{sec:ff}
%

Here we briefly outline the main features of our method and we refer
the reader to our earlier papers for more details
\cite{Boyle:2007qe,Boyle:2007wg,Boyle:2008yd}.

The matrix element of the vector current between initial and final
state pseudoscalar mesons $P_i$ and $P_f$, is in general decomposed
into two invariant form factors:
\begin{equation}
  \langle {P_f(p_f)}|V_{\mu}| {P_i(p_i)}\rangle =
  f^+_{P_iP_f}(q^2)(p_i+p_f)_{\mu}+f^-_{P_iP_f}(q^2)(p_i-p_f)_{\mu},
\label{eq:me}
\end{equation}
where $q^2=-Q^2=(p_i-p_f)^2$.
For $K \rightarrow \pi$, $V_{\mu} = \bar{s}\gamma_{\mu}u$, $P_i = K$
and $P_f = \pi$.
For $\pi \rightarrow \pi$, $V_{\mu} =
\frac{2}{3}\bar{u}\gamma_{\mu}u-\frac{1}{3}\bar{d}\gamma_{\mu}d$, $P_i
= P_f= \pi$ and from vector current conservation,
$f^-_{\pi\pi}(q^2)=0$.
The form factors $f^+_{P_iP_f}(q^2)$ and $f^-_{P_iP_f}(q^2)$ contain
the non-perturbative QCD effects and hence are ideally suited for a
determination in lattice QCD.

In a finite volume with spatial extent $L$ and periodic boundary
conditions for the quark fields, momenta are discretised in units of
$2\pi/L$.
As a result, the minimum non-zero value of $Q^2$ for the pion form
factor in our simulation is $q^2_{\rm min}\approx -0.15\ {\rm GeV}^2$,
while for the $K\to\pi$ form factor
\be
q^2=(E_K(\vec{p}_i)-E_\pi(\vec{p}_f))^2 - (\vec{p}_i - \vec{p}_f)^2\ .
\ee
For $\vec{p}_i=0$ and $2\pi/L$ with $\vec{p}_f=0$, we have $q^2\approx
0.06\ {\rm GeV}^2$ and $-0.05\ {\rm GeV}^2$, respectively,
presenting the need for an interpolation in order to extract the
result of the form factor, $f_{K\pi}^+$, at $q^2=0$.

In order to reach small momentum transfers for the pion form factor
and $q^2=0$ for the $K\to\pi$ form factors, we use partially twisted
boundary conditions \cite{Sachrajda:2004mi,Bedaque:2004ax}, combining
gauge field configurations generated with sea quarks obeying periodic
boundary conditions with valence quarks with twisted boundary
conditions
\cite{Sachrajda:2004mi,Bedaque:2004ax,Bedaque:2004kc,deDivitiis:2004kq,Tiburzi:2005hg,Flynn:2005in,Guadagnoli:2005be}.
The valence quarks, $q$, satisfy
\be
q(x_k+L) = e^{i\theta_k}q(x_k),\qquad (k=1,2,3)\,,
\ee
where $\vec{\theta}$ is the twisting angle.
\begin{center}\begin{picture}(120,60)(-60,-30) \ArrowLine(-50,0)(-25,0)
\ArrowLine(25,0)(50,0)\Oval(0,0)(12,25)(0)
\GCirc(-25,0){3}{0.5}\GCirc(25,0){3}{0.5} \GCirc(0,12){3}{0.5}
\Text(-19,12)[b]{$q_2$}\Text(19,12)[b]{$q_1$}
\Text(0,-15)[t]{$q_3$}\Text(0,17)[b]{$V_\mu$}\Text(-54,0)[r]{$P_i$}
\Text(54,0)[l]{$P_f$}\ArrowLine(0.5,-12)(-0.5,-12)
\end{picture}\end{center}
Our method is decribed in detail in \cite{Boyle:2007wg,Boyle:2008yd}
and proceeds by setting $\vec{\theta}=0$ for the spectator quark,
denoted by $q_3$ in the above diagram.
We are then able to vary the twisting angles, $\vec{\theta}_i$ and
$\vec{\theta_f}$, of the quarks before $(q_2)$ and after $(q_1)$ the
insertion of the current, respectively.
The momentum transfer between the initial and final state mesons is
now
\be
q^2=(E_i(\vec{p}_i,\vec{\theta}_i)-E_f(\vec{p}_f,\vec{\theta}_f))^2 -
((\vec{p}_i+\vec{\theta}_i/L) - (\vec{p}_f+\vec{\theta}_f/L))^2\ ,
\ee
where
$E(\vec{p},\vec{\theta})=\sqrt{m^2+(\vec{p}+\vec{\theta}/L)^2}$.
Hence it is possible to choose $\vec{\theta}_i$ and $\vec{\theta}_f$
such that $q^2=0$, which from now on we refer to as $\vec{\theta}_K$
and $\vec{\theta}_\pi$ for when we twist a quark in the Kaon and Pion,
respectively.

In order to extract the matrix elements (\ref{eq:me}) from a lattice
simulation, we consider ratios of three- and two-point correlation
functions.
For the pion form factor, we consider the ratios given in Eqs.~(3.4)
and (3.5) in \cite{Boyle:2008yd}, while for the $K\to\pi$ form
factors, we consider the following ratios
\begin{equation}\label{eq:ratios}
\begin{array}{rcl}
R_{1,\,P_iP_f}(\vec{p}_i,\vec{p}_f)&{=}&
 4\sqrt{E_i E_f}\, \sqrt{\frac
 {C_{P_iP_f}(t,\vec p_i,\vec p_f)\,C_{P_fP_i}(t,\vec p_f,\vec p_i)}
 {C_{P_i}(t_{\rm sink},\vec p_i)\,C_{P_f}(t_{\rm sink},\vec p_f)}},
 \\[4mm]
R_{3,\,P_iP_f}(\vec{p}_i,\vec{p}_f)&=&
 4{\sqrt{E_i E_f}}\,
\frac{C_{P_iP_f}(t,\vec p_i,\vec p_{f})}{C_{P_f}(t_{\rm sink},\vec p_f)}\,
    \sqrt{
    \frac{C_{P_i}(t_{\rm sink}-t,\vec p_i)\,C_{P_f}(t,\vec p_f)\,C_{P_f}(t_{\rm sink},\vec p_f)}
    {C_{P_f}(t_{\rm sink}-t,\vec p_f)\,C_{P_i}(t,\vec p_i)\,C_{P_i}(t_{\rm sink},\vec
    p_i)}}\,.
\end{array}
\end{equation}

We deviate slightly from the method outlined in \cite{Boyle:2007wg}
for extracting $f_{K\pi}^0(0)$ from the ratios.
Previously we considered only the time-component of the vector current
and solved for $f_{K\pi}^0(0)=f_{K\pi}^+(0)$ via the linear combination
\begin{equation}\label{eq:lin_comb}
f_{K\pi}^0(0)=\frac{ R_{\alpha,K\pi}(\vec{p}_K,\vec{0})(m_K-E_\pi)
- R_{\alpha,K\pi}(\vec{0},\vec{p}_\pi)(E_K-m_\pi) }{
(E_K+m_\pi)(m_K-E_\pi)-(m_K+E_\pi)(E_K-m_\pi)
}\qquad(\alpha=1,2,3)\,.
\end{equation}
This, however, is just one of many expressions that can be obtained
when we solve the system of simultaneous equations that are obtained
when we consider all components of the vector current, $V_\mu$, rather
than just $V_4$ that was considered in \cite{Boyle:2007wg}
\begin{eqnarray}
R_{\alpha,K\pi}(\vec{\theta}_K,\vec{0},V_4) &=& 
    f_{K\pi}^+(0)\,(E_K+m_\pi) + f_{K\pi}^-(0)\,(E_K-m_\pi)\nonumber\\
R_{\alpha,K\pi}(\vec{0},\vec{\theta}_\pi,V_4) &=& 
    f_{K\pi}^+(0)\,(m_K+E_\pi) + f_{K\pi}^-(0)\,(m_K-E_\pi)\nonumber\\
R_{\alpha,K\pi}(\vec{\theta}_K,\vec{0},V_i) &=& 
    f_{K\pi}^+(0)\,\theta_{K,i} + f_{K\pi}^-(0)\,\theta_{K,i}\nonumber\\
R_{\alpha,K\pi}(\vec{0},\vec{\theta}_\pi,V_i) &=& 
    f_{K\pi}^+(0)\,\theta_{\pi,i} - f_{K\pi}^-(0)\,\theta_{\pi,i}\ .
\end{eqnarray}
We can now proceed to solve this overdetermined system of equations
via $\chi^2$ minimisation.

%
\section{Pion form factor results}
%

\begin{figure}
\begin{tabular}{lcr}
\hspace*{-1mm}
	\psfrag{xlabel}[c][b][1][0]{\small $Q^2[\gev^2]$}
\psfrag{ylabel}[c][t][1][0]{\small $f^{\pi\pi}(q^2)$}
\includegraphics[width=0.35\textwidth,angle=-90]{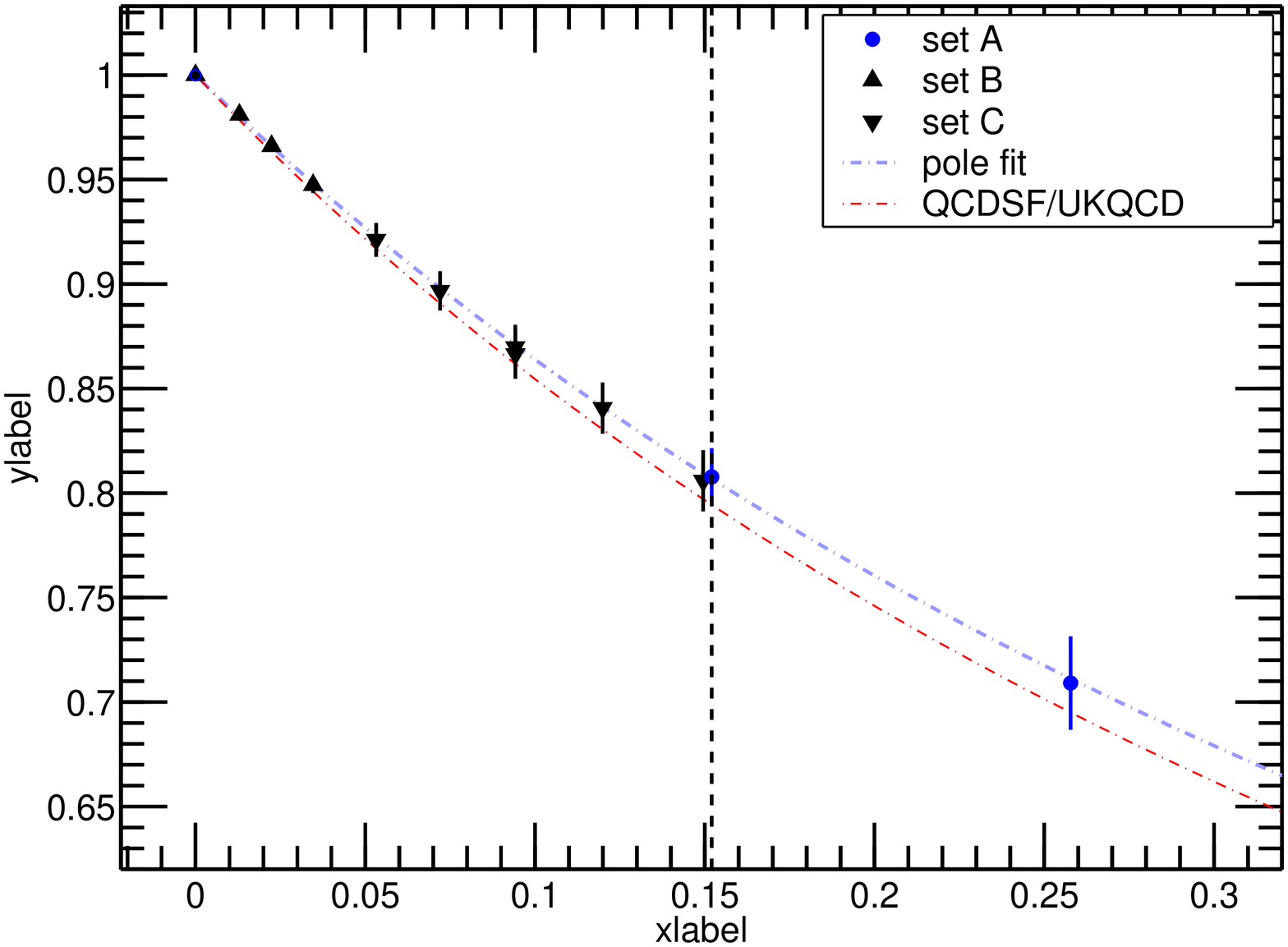} &&
\hspace*{-6mm}
\psfrag{xlabel}[c][bc][1][0]{\small $Q^2[\gev^2]$}
\psfrag{ylabel}[c][t][1][0]{\small $f^{\pi\pi}(q^2)$}
\psfrag{exp}[l][lc][1][0]{\tiny experimental data NA7}
\psfrag{330MeV}[l][lc][1][0]{\tiny lattice data for $m_\pi=330\mev$}
\psfrag{NLO330MeV}[l][ll][1][0]{\tiny $\SU(2)$ NLO lattice-fit; $m_\pi=330\mev$}
\psfrag{NLO139.57MeV}[l][ll][1][0]{\tiny $\SU(2)$ NLO lattice-fit;
		$m_\pi=139.57\mev$}
\psfrag{444OOOOOOOOOOOOOOOOOOOO}[l][lc][1][0]{\tiny $1+\frac 16
		\langle r^2_\pi\rangle^{\rm PDG}Q^2$}
	\includegraphics[width=0.35\textwidth,angle=-90]{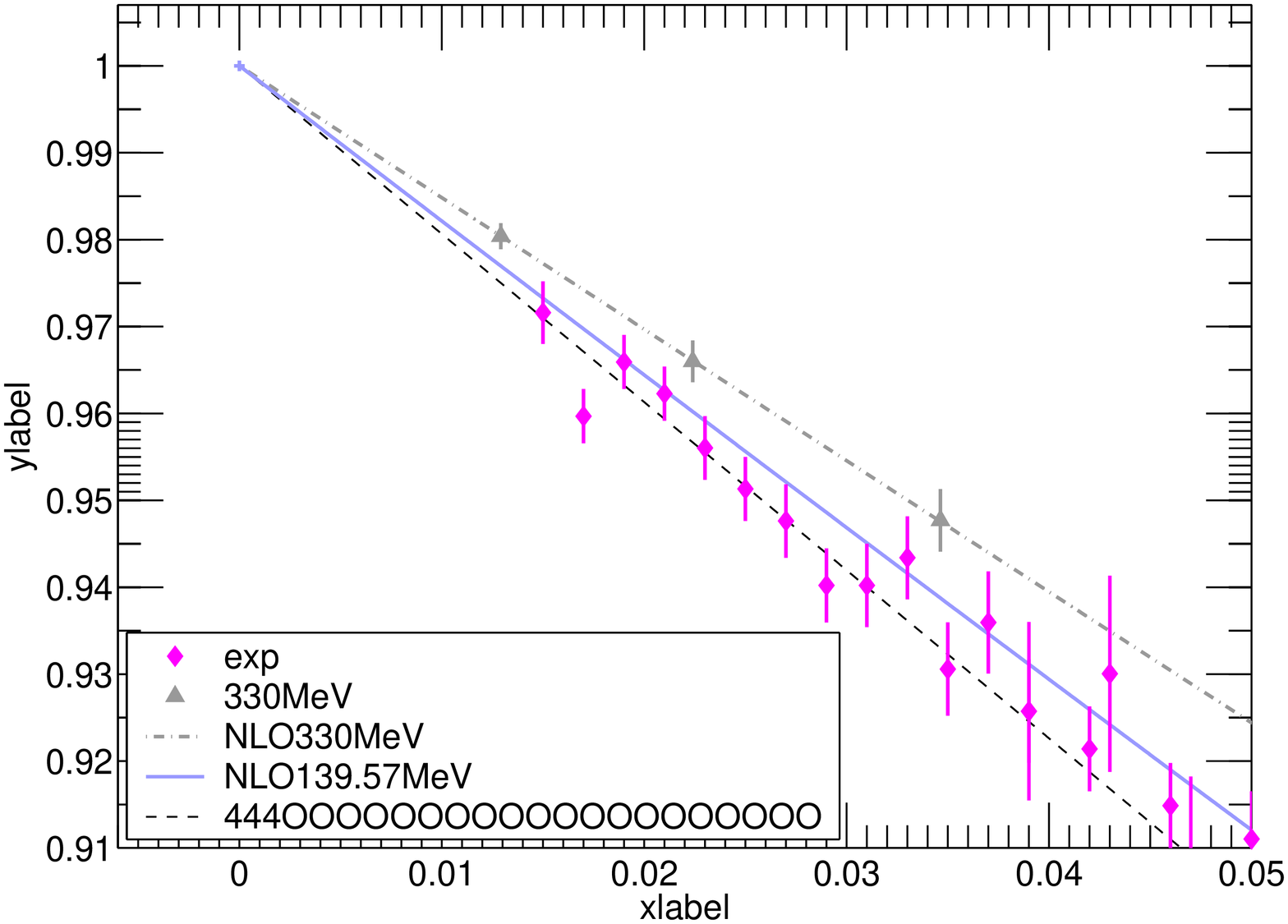}	
	\end{tabular}
\caption{$f^{\pi\pi}(q^2)$ from a $24^3\times 64$ lattice with
  $m_\pi=330$~MeV using partially twisted bc's.}
\label{fig:fpipi}
\end{figure}

In Fig.~\ref{fig:fpipi} we show our results for the form factor
$f^{\pi\pi}(q^2)$ for a pion with $m_\pi=330\mev$ for a range of
values of $q^2$ both using periodic bc's and partially twisted bc's
(set A and sets B\&C respectively in the left plot of figure).
The vertical dashed line indicates the smallest momentum transfer
available on this lattice with periodic bc's.
The (blue) dashed line is the result of a pole-dominance fit to our
data points, while the (red) dot-dashed curve is obtained from the
result of QCDSF \cite{Brommel:2006ww} evaluated at $m_\pi=330$~MeV.

On the right of Fig.~\ref{fig:fpipi} we have a zoom into the low $Q^2=-q^2$
region. The triangles are our lattice data points for a pion with
$m_\pi=330\mev$, and the magenta diamonds are experimental data points
for the physical pion.

Because our values of $Q^2$ are very small, we apply NLO chiral
perturbation theory (ChPT).
In NLO ChPT, the pion form factor depends only on a single low energy
constant (LEC) ($L_9^r$ for SU(3), or $l_6^r$ for SU(2))
\begin{eqnarray}
f^{\pi\pi}_{\SU(2),\mathrm{NLO}}(q^2) &=&
 1+\frac1{f^2}\left[
   -2l_6^r \,q^2 + 4\tilde{\mathcal{H}}(m_\pi^2,q^2,\mu^2)\right]
\label{eq:fpipiSU2}\\
f^{\pi\pi}_{\SU(3),\mathrm{NLO}}(q^2) &=&
 1+\frac1{f_0^2}\left[
   4L_9^r \,q^2 + 4\tilde{\mathcal{H}}(m_\pi^2,q^2,\mu^2)
             + 2\tilde{\mathcal{H}}(m_K^2,q^2,\mu^2)\right]
\label{eq:fpipiSU3}
\end{eqnarray}
where
\begin{equation}
\tilde{\mathcal{H}}(m^2,q^2,\mu^2) =
 \frac{m^2 H(q^2/m^2)}{32\pi^2} -
  \frac{q^2}{192\pi^2}\log\frac{m^2}{\mu^2}
\end{equation}
and
\begin{equation}\label{eq:Hdef}
H(x) \equiv -\frac43 + \frac5{18}x -
 \frac{(x-4)}6 \sqrt{\frac{x-4}x}
 \log\left(\frac{\sqrt{(x-4)/x}\,+1}{\sqrt{(x-4)/x}\,-1}\right)
\end{equation}
with $H(x) = -x/6 + O(x^{3/2})$ for small $x$.
Provided our pion mass is light enough, we can use the $q^2$
dependence of $f^{\pi\pi}(q^2)$ to extract this LEC.
The grey dashed curve on the right hand of Fig.~\ref{fig:fpipi} shows
our SU(2) fit to the $m_\pi=330\mev$ pion form factor data.
 
Once the LEC is determined from this fit, we insert the physical pion
mass in (\ref{eq:fpipiSU2}) to obtain the solid blue curve. 
In addition we also represent the PDG world average \cite{Yao:2006px}
for the charge radius using the black dashed line.
Our best estimate for the pion charge radius comes from the SU(2) NLO
ChPT fit to the three lowest $Q^2$ points and is
\begin{equation} 
\rpisq=\rpisqphyslong \,\rm{fm^2}\ .
\end{equation}
The fact that our result is in agreement with experiment,
$\rpisq=0.452(11) \,\rm{fm^2}$ \cite{Yao:2006px}, gives us confidence
that we are in a regime where chiral perturbation theory is
applicable.

%
\section{$K_{l3}$ form factor results}
%

As explained in Sec.~\ref{sec:ff}, we calculate the $K \rightarrow
\pi$ form factor directly at $q^2=0$ by setting the Kaon and Pion in
turn to be at rest, while twisting the other one such that $q^2=0$.
We refer to these twist angles as $\theta_\pi$ and $\theta_K$,
respectively.
We then get the following equations:
\begin{eqnarray}
\langle K(p_K)|V_\mu| \pi(0)\rangle &=&
    f_{K\pi}^+(0)p_{K,\mu} -
    f_{K\pi}^-(0)p_{K,_\mu} \nonumber\\
\langle K(0)|V_\mu| \pi(p_{\pi})\rangle &=&
    f_{K\pi}^+(0)p_{\pi,\mu} +
    f_{K\pi}^-(0)p_{\pi,_\mu}
\label{eq:simeq}
\end{eqnarray}
\begin{wrapfigure}{r}{0.48\textwidth}
\includegraphics[width=0.5\textwidth]{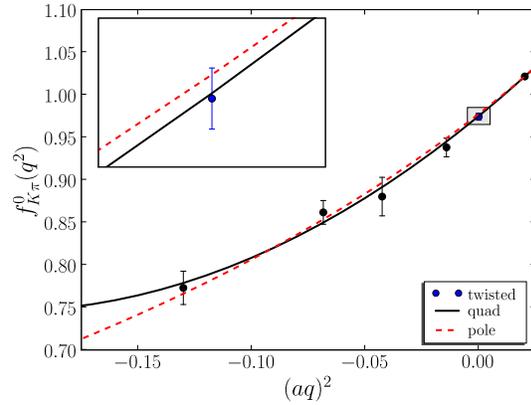}
\caption{$K_{\ell 3}$ form factor, $f_{K\pi}^0(q^2)$, evaluated at $q^2=0$
  directly using twisted boundary conditions. Results are compared
  with data at $q^2\ne 0$ and fits from \cite{Boyle:2007qe}}
\label{fig:kl3}
\end{wrapfigure}
By simply solving the simultaneous equations for each of the
$\mu$ components separately we find that the errors in $f_{K\pi}^+(0)$ and
$f_{K\pi}^-(0)$, are much larger than the errors in the matrix elements.
We have managed to circumvent this by looking at all the $\mu$
components simultaneously, and then performing a $\chi^2$ minimisation
on the overdetermined system of equations to find the values of
$f_{K\pi}^+(0)$ and $f_{K\pi}^-(0)$ that best fit the equations.

To obtain the matrix elements (\ref{eq:simeq}), we consider different
combinations of $R_1$ and $R_3$ (\ref{eq:ratios}).
We find that all combinations lead to consistent results, with the
best combination being that we use $R_3$ for all matrix elements
except for the case where the pion is twisted and we are considering
the $4^{\rm th}$ component of the vector current.
Using this set up, we obtain our preliminary results for
$f^+_{K\pi}(0)$ and $f^-_{K\pi}(0)$ (for a pion mass of
$m_{\pi}=330\mev$)
\begin{equation}  
f^+_{K\pi}(0) = 0.9742(41)\,,\quad f^-_{K\pi}(0)=-0.113(12)\ .
\end{equation}
Our result for $f^+_{K\pi}(0)=f^0_{K\pi}(0)$ is shown in
Fig.~\ref{fig:kl3} where we compare with the previous determinations
in \cite{Boyle:2007qe} which used pole $f^+_{pole}(0) = 0.9774(35)$
and quadratic $f^+_{quad}(0) =0.9749(59)$ functions to
interpolate between $q^2_{max}$ and negative values of $q^2$.
In our previous result, $f_{K\pi}^+(0) = 0.9644(33)(34)(14)$,
these were combined, taking a systematic error of (34) for the model
dependence.  
This contribution to the error has been eliminated in our new
calculation.

We conclude that using partially twisted bc's for the $K_{\ell 3}$ form
factor, is an improvement on the conventional method as it removes a
source of systematic error, while keeping comparable statistical
errors.
Another source of systematic error in our result in
\cite{Boyle:2007qe} is due to the slight difference between our
simulated strange quark mass ($am_s+am_{\rm res}\simeq 0.043$) and the
physical strange quark ($am_s+am_{\rm res}\simeq 0.037$)
\cite{Allton:2008pn}, and we are in the process of determining the
effect this has on our result through a simulation with a partially
quenched strange quark mass of $am_s+am_{\rm res}\simeq 0.033$.
We also plan to combine our results with the latest expressions from
chiral perturbation theory \cite{Flynn:2008tg}.

%
\section*{Acknowledgements}
%

We thank our colleagues in RBC and UKQCD within whose programme this
calculation was performed. 
We thank the QCDOC design team for developing the QCDOC machine and
its software.
This development and the computers used in this calculation were
funded by the U.S.DOE grant DE-FG02-92ER40699, PPARC JIF grant
PPA/J/S/1998/0075620 and by RIKEN. 

We thank the University of Edinburgh, PPARC, RIKEN, BNL and the
U.S. DOE for providing the QCDOC facilities used in this calculation.
We are very grateful to the Engineering and Physical Sciences Research
Council (EPSRC) for a substantial allocation of time on HECToR under
the Early User initiative. 
We thank Arthur Trew, Stephen Booth and other EPCC HECToR staff for
assistance and EPCC for computer time and assistance on BlueGene/L.

JMF, AJ, HPdL and CTS acknowledge support from STFC Grant PP/D000211/1
and from EU contract MRTN-CT-2006-035482 (Flavianet). PAB, CK, CMM
acknowledge support from STFC grant PP/D000238/1. JMZ acknowledges
support from STFC Grant PP/F009658/1.

\end{document}